\documentclass[12pt,a4paper,reqno]{amsart}
\usepackage{amssymb,amsmath}
\usepackage{hyperref}
\usepackage{setspace}
\usepackage{graphicx}

\theoremstyle{definition}

\begin{document}

\begin{center}
{\Large Multi-dimensional Black-Scholes-Merton to Hamilton-Jacobi}

\bigskip

\textbf{Muhammad Naqeeb}

Department of Mathematics, Quaid-i-Azam University, Islamabad-44000, Pakistan.

Email; mnaqeem@math.qau.edu.pk

\bigskip
\bigskip

\textbf{Amjad Hussain}

Department of Mathematics, Quaid-i-Azam University, Islamabad-44000, Pakistan.

Email; a.hussain@qau.edu.pk

\bigskip

\end{center}

\begin{center}
{\small Abstract}
\end{center}
   The first widely used financial model to price the evolution of options has footprints almost in every field. The aim of this paper is to derive the Black-Scholes-Merton formula of multiple options, generally for an n-dimensional assets and its links to Hamilton-Jacobi equation of mechanics with solution of black-Scholes equation in the metric of Banach space.\par
keywords: Multi-dimensional Black-Scholes-Merton, Hamilton-Jacobi European call option, Banach space.

\begin{center}
{\Large Introduction}
\end{center}
Dynamical systems, deterministic or stochastic are interlinked with Mathematical Finance. In Dynamical Mathematical Finance we examine financial models by mean of dynamical systems. Dynamical systems defined over multiple independent variables categorized as multi-dimensional systems. For Black-Scholes model of Mathematical Finance one or multiple underlying assets could be consider. For one underlying asset the model is,
\begin{equation}
\frac{\partial v}{\partial t} +rs_1\frac{\partial^2 v}{\partial s_1^2}+\frac{1}{2}\sigma^2s_1^2\frac{\partial^2 v}{\partial s_1^2}-rv=0
\end{equation}
For an n-underlying assets or multi-assets the formula, which is derived in section one, is,
\begin{equation}\begin{aligned}&
\frac{\partial v}{\partial t} + \frac{1}{2}\sigma_1^2 s_1^2 \frac{\partial^2 v}{\partial s_1^2}+ \frac{1}{2}\sigma_2^2 s_2^2 \frac{\partial^2 v}{\partial s_2^2}+...+\frac{1}{2}\sigma_n^2 s_n^2 \frac{\partial^2 v}{\partial s_n^2}+\\& \frac{\partial v}{\partial s_1}ds_1+\frac{\partial v}{\partial s_2}ds_2+...+\frac{\partial v}{\partial s_n}ds_n+
rs_1\frac{\partial v}{\partial s_1}+
rs_2\frac{\partial v}{\partial s_2}+...+rs_n\frac{\partial v}{\partial s_n}+\\&\rho_{1n}\sigma_1 \sigma_n \frac{\partial^2 v}{\partial s_1s_n}+
\rho_{2n}\sigma_2 \sigma_n \frac{\partial^2 v}{\partial s_2s_n}+...+\rho_{n-1n}\sigma_2 \sigma_n \frac{\partial^2 v}{\partial s_n-1_n}=0\end{aligned}
\end{equation}
The Hamilton-Jacobi equation, an alternate formulation of Classical Mechanics and necessary condition for describing extremal geometry in generalization of problems from the calculus of variations. For one spatial or generalized coordinate the equation is give as,
\begin{equation}
\frac{\partial U}{\partial t}+H(q_1, \frac{\partial U}{\partial q_1},)
\end{equation}
For an n-generalized coordinates or spatial variables the above equation after extension is give as,
\begin{equation}
\frac{\partial U}{\partial t}+H(q_1,q_2,q_3,...q_n, \frac{\partial U}{\partial q_1}, \frac{\partial U}{\partial q_2},\frac{\partial U}{\partial q_3},...,\frac{\partial U}{\partial q_n} )
\end{equation}
Rodrigo and Galvez in \cite{BSTH} presented their work on relationships between equation (1) and (3) for one spatial variable considered as an underlying asset. In present work, the relationships between equation (2) and (4) are studied and explained, Which are multi-dimensional and multi-asset models. The work is presented systematically by dividing it in different sections.\par
The first one is to derive and analyze Black-Scholes Merton having an n-underlying assets, the second section is to dicuss Hamilton-Jacobi equation in multi-variable calculus. Moving to third major section that explains relationships between Black-Bcholes-Merton and Hamilton-Jacobi equation of mechancis with concluding remarks and suggestions in fourth section

\section{black scholes equation} From everyday markets to the implications of theory of Relativity in financial mathematics  \cite{BT}. This formula beyond the achievement of Noble price in 1997 for Scholes and Merton \cite{M1} has impressed every field. To derive having an n-underlying assets starting from multiple stochastic process

\begin{equation*}
\frac{ds_1}{s_1}=\mu_1dt+\sigma_1dw_1    \quad , \quad
\frac{ds_2}{s_2}=\mu_2dt+\sigma_2dw_2
\end{equation*}
\begin{equation*}\begin{aligned}
\frac{ds_n}{s_n}=\mu_ndt+\sigma_ndw_n  \end{aligned}
\end{equation*}
\quad
where $dw_n ; n=1,2,...$ is Brownian motion.
\begin{equation*}
E(dw_i)=0 ;  E(dw_i^2)=dt
\end{equation*}
By definition of random walk or wiener process \cite{BTZ} \par
Correlation factor which appears to be zero in case of one underlying assest here it will be in relation with each two distinct random walks to determine their strength from positive one, zero or negative one
  correlation factor having range -1 to +1 is quite different from risk free interest rate ranging from
  0 to +1.
  These factors are vital to consider.
  \begin{equation*}
  \Pi=v-\triangle_1s_1+\triangle_2s_2+...+\triangle_ns_n
  \end{equation*}
  \begin{equation*}
  d\Pi=dv-\triangle_1ds_1+\triangle_2ds_2+...+\triangle_nds_n
  \end{equation*}
\begin{text} Now, the n-dimensional ito's lemma is given as
\end{text}
\begin{equation}\begin{aligned}
dv&=\frac{\partial v}{\partial t} + \frac{1}{2}\sigma_1^2  \frac{\partial^2 v}{\partial s_1^2}+ \frac{1}{2}\sigma_2^2 \frac{\partial^2 v}{\partial                  s_2^2}+...+\frac{1}{2}\sigma_n^2  \frac{\partial^2 v}{\partial s_n^2}\\
&+ \frac{\partial v}{\partial s_1}ds_1+\frac{\partial v}{\partial s_2}ds_2+...+\frac{\partial v}{\partial s_n}ds_n+
\rho_{1n}\sigma_1 \sigma_n \frac{\partial^2 v}{\partial s_1s_n}\end{aligned}
\end{equation}
\begin{equation*}
\rho_{2n}\sigma_2 \sigma_n \frac{\partial^2 v}{\partial s_2s_n}+...+\rho_{n-1n}\sigma_2 \sigma_n \frac{\partial^2 v}{\partial s_n-1_n}
\end{equation*}
To eliminate the risk terms from above n ito's lemma
\begin{equation*}
\triangle_1=\frac{\partial v}{\partial s_1}  ,   \triangle_2= \frac{\partial v}{\partial s_2}  ,...,  \triangle_n=\frac{\partial v}{\partial s_n}
\end{equation*}
\begin{equation}\begin{aligned}
dv&=\frac{\partial v}{\partial t} + \frac{1}{2}\sigma_1^2 s_1^2 \frac{\partial^2 v}{\partial s_1^2}+
 \frac{1}{2}\sigma_2^2 s_2^2 \frac{\partial^2 v}{\partial s_2^2}+...+\frac{1}{2}\sigma_n^2 s_n^2 \frac{\partial^2 v}{\partial s_n^2}+\\& \frac{\partial v}{\partial s_1}ds_1+\frac{\partial v}{\partial s_2}ds_2+...+\frac{\partial v}{\partial s_n}ds_n+
\rho_{1n}\sigma_1 \sigma_n \frac{\partial^2 v}{\partial s_1s_n}\end{aligned}
\end{equation}
\begin{equation*}
\rho_{2n}\sigma_2 \sigma_n \frac{\partial^2 v}{\partial s_2s_n}+...+\rho_{n-1n}\sigma_2 \sigma_n \frac{\partial^2 v}{\partial s_n-1_n}
\end{equation*}
\begin{equation*}\begin{aligned}&
d\Pi=\text{riskless return rate would be}\quad\\
&d\Pi=r\Pi=r\{v-\frac{\partial v}{\partial s_1}s_1-\frac{\partial v}{\partial s_2}s_2-...-\frac{\partial v}{\partial s_n}s_n\}dt\end{aligned}
\end{equation*}
and finally we conclude Black-Scholes equation which is
\begin{equation}\begin{aligned}&
\frac{\partial v}{\partial t} + \frac{1}{2}\sigma_1^2 s_1^2 \frac{\partial^2 v}{\partial s_1^2}+ \frac{1}{2}\sigma_2^2 s_2^2 \frac{\partial^2 v}{\partial s_2^2}+...+\frac{1}{2}\sigma_n^2 s_n^2 \frac{\partial^2 v}{\partial s_n^2}+\\& \frac{\partial v}{\partial s_1}ds_1+\frac{\partial v}{\partial s_2}ds_2+...+\frac{\partial v}{\partial s_n}ds_n+
rs_1\frac{\partial v}{\partial s_1}+
rs_2\frac{\partial v}{\partial s_2}+...+rs_n\frac{\partial v}{\partial s_n}+\\&\rho_{1n}\sigma_1 \sigma_n \frac{\partial^2 v}{\partial s_1s_n}+
\rho_{2n}\sigma_2 \sigma_n \frac{\partial^2 v}{\partial s_2s_n}+...+\rho_{n-1n}\sigma_2 \sigma_n \frac{\partial^2 v}{\partial s_n-1_n}=0\end{aligned}
\end{equation}
This is the required derived black scholes equation having multiple of n underlying assets, which affectively differ from
black scholes equation in one underlying asset especially in appearance of correlation factors terms
having their intensity defined below
\begin{equation*}
(dw_1dw_n)=\rho_{1n},   \{dw_2dw_n\}=\rho_{1n}, ... ,\{dw_n-1dw_n\}=\rho_{n-1n}
\end{equation*}
These are correlation coefficients for an n underlying assets, after discussing their terms of correlation coefficients. here number of terms in an equation could be random but still predicted which is Mathematically healthy in number theory like
first equation having 4 terms the next one having 4+3=7, and next one having 7+4=11 terms, then 11+5=16, then 16+6=22 terms and so on. s special kind of series at predicts terms containing in Black-Scholes equation having multiple assets
\subsection {conditions and solutions:} Considering space which here best suited is n-dimensional Euclidean space, whose discussion will follow later in third section , which will become the Banach space in which solution could be defined by introducing special metric is vital, to consider v the solution variable in n dimensional space.

Considering space $ \mathbb R^n$ having solution of the form
\begin{equation*}\begin{aligned}&
v(s_1,s_2,s_3,...,s_n )=f(s_1,s_2,s_3,...,s_n ,t)\end{aligned}
\end{equation*}
The Initial condition for an n underlying asset for an European call option would be
\begin{equation*}
v\left(s_1,s_2,s_3,...,s_n, T\right)=max\{max\left(s_1,s_2,s_3,...,s_n)-k,0\right\}
\end{equation*}
While K is an expiration or exercise date on which option has to be exercised.\par
The above model is of Multi dimensional Black scholes equation with European conditions, Although our discussion proceeds considering European call option, The discussion could be expanded by taking conditions on European put options, American put and call options, Basket options and for exchange options.
At the expiry date 'K' the solution, By reducing Multi-dimensional Black-scholes to diffusion equation in n dimensions \cite{BT} possessing 'N' a standard normal cumulative distribution function is
\begin{equation}\begin{aligned}&
V\left(s_1,s_2,s_3,...,s_n ,T\right)=s_1N\left(\phi_{1},p_1\right)+s_2N\left(\phi_{2},p_2\right)+\\
&s_3N\left(\phi_{3},p_3\right)...+s_N\left(\phi_{n},p_n\right)N-Ke^{-rT}\{1-\{N\left(\phi_{1}'\right)\\
&+ N\left(\phi_{2}'\right)+N\left(\phi_{3}'\right)+...+N\left(\phi_{n}'\right)\}\}\end{aligned}
\end{equation}
\section{Hamilton jacobi differential equation and its solutions}
\begin{center}
Hamilton jacobi
\end{center}
In dynamical systems the non linear Hamilton-Jacobi equation could be used to derive equations of motion. The Hamilton-Jacobi equation in (n+1) variables
\begin{equation*}
q_1,q_2,q_3,...q_n,t \quad is
\end{equation*}
\begin{equation}
\frac{\partial U}{\partial t}+H(q_1,q_2,q_3,...q_n, \frac{\partial U}{\partial q_1}, \frac{\partial U}{\partial q_2},\frac{\partial U}{\partial q_3},...,\frac{\partial U}{\partial q_n})
\end{equation}
The Black-Scholes model and Hamilton-Jacobi, Both models are liked to each other in terms of Initial conditions, Spaces, Energy relations, Domains, Spatial variables. Employing mathematical techniques the dynamical systems to some extent could be linked to models in finance and economics, With the help of related concepts and different fields.  \cite{BSTH}
\par
The Hamilton-Jacobi in Variational calculus is a cauchy problem\cite{EVAN} which is represented by
\begin{equation*}
\frac{\partial U}{\partial t}+H(q_1,q_2,q_3,...q_n, \frac{\partial U}{\partial q_1}, \frac{\partial U}{\partial q_2},\frac{\partial U}{\partial q_3},...,\frac{\partial U}{\partial q_n})=0
\end{equation*}
\begin{equation*}\begin{aligned}&
\text{where} \quad R^n\times(0,\infty),
\text{with Initial condition}\quad U=(g_1,g_2,...g_n),\\
&\text{and} \quad (g_1,g_2,...g_n)\in \mathbb{R^n}\times (t=0)\end{aligned}
\end{equation*}
\begin{equation*}
\text{Here we have}\quad U:\mathbb{R^n}\times(0,\infty)\rightarrow\mathbb{ R}\quad \text{and Hamiltonian defined by}
\end{equation*}
\begin{equation*}
H:R^n\rightarrow R,\quad \text{where}\quad (g_1,g_2,...g_n):R^n\rightarrow R
\end{equation*}
\subsection{solution of Hamilton-Jacobi equation}
The solution of the Hamilton-Jacobi equation in n-dimensions, Or In variational calculus will be:
\begin{equation*}
Let \quad L:R^n\rightarrow R,\quad \text{name it Lagrangian function satisfying conditions}
\end{equation*}
\begin{equation*}
The \quad mapping\quad q\longmapsto L(q)\quad is \quad convex
\end{equation*}
\begin{equation*}
lim \frac{L(q)}{|q|}=\infty \quad \quad as \quad q\rightarrow \infty
\end{equation*}
The convexity implies L is continuous lagrangian function. We can obtain Hamiltonian from Lagrangian by the Legendre transformation of L. Hamiltonian and Lagrangian are both dual complex functions, interpreting, H=L, The form of Hamiltonian and Lagrangian is almost same. To discuss solution we try to minimize the action introduced as
\begin{equation*}
I\{W(.)\}=\int_{0}^{t}L(\dot w(s)ds \quad  over \quad W:[0,t]\rightarrow R^n
\end{equation*}
\begin{equation*}
where \quad W(.)=\{W^1(.),W^2(.)W^3(.),...,W^1(.)\}
\end{equation*}
Modifying the above equation to include the function g evaluated at W(0)
\begin{equation*}
I\{W(.)\}=\int_{0}^{t}L\{\dot W(s)ds\}+g\{w(0)\}
\end{equation*}
The solution of the Hamilton-Jacobi equation in terms of Variational principle entailing this modified action is
\begin{equation*}
U(x,t)=\inf\Bigg\{\int_{0}^{t}L\{\dot W(s)\}ds+g(y)|  W(0)=y,\quad W(t)=x
\end{equation*}
\begin{equation*}
\text{Whereas infimum is taken over all}\quad W(.)\in C^1 \quad\text{ with} \quad W(t)=x.
\end{equation*}
H is smooth and Convex, While $(g_1,g_2,...g_n)$ must be Lipschitz continuous. The above minimization problem could be simplified as
\begin{equation*}
U(x,t)=min_{y\in R^n} \Bigg\{ tL(\frac {x-y}{t})+ g(y)
\end{equation*}
The expression on the right hand side is called Hopf-Lax formula. This formula is a unique weak solution of the initial value problem for the Hamilton-Jacobi equation. While in variational analysis the solution of the cauchy problem 'U' is defined in the metric of Banach space, L is invariant\cite{RUND} while H must be of the class of second differential functions, Hamilton-Jacobi equation is an evolution equation
\section{On Relations of Multi-dimensional Black-Scholes-Merton and Hamilton-Jacobi}
\subsection{} Both are non-linear evolution equations with same degree and order consisting of Hamiltonian and Lagrangian
\subsection{}
Multi-dimensional black-Scholes-Merton equation having solution of the form
\begin{equation*}
V=F\left(s_1,s_2,s_3,...,s_n ,T\right)
\end{equation*}
If we express Black-Scholes-Merton in the form
\begin{equation}
\begin{aligned}&
\frac{\partial v}{\partial t} + \frac{1}{2}\sigma_1^2 s_1^2 \frac{\partial^2 v}{\partial s_1^2}+ \frac{1}{2}\sigma_2^2 s_2^2 \frac{\partial^2 v}{\partial s_2^2}+...+\frac{1}{2}\sigma_n^2 s_n^2 \frac{\partial^2 v}{\partial s_n^2}+\\& \frac{\partial v}{\partial s_1}ds_1+\frac{\partial v}{\partial s_2}ds_2+...+\frac{\partial v}{\partial s_n}ds_n+
rs_1\frac{\partial v}{\partial s_1}+\\&
rs_2\frac{\partial v}{\partial s_2}+...+rs_n\frac{\partial v}{\partial s_n}+
\rho_{1n}\sigma_1 \sigma_n \frac{\partial^2 v}{\partial s_1s_n}-rv=0\end{aligned}
\end{equation}
for n underlying assets  we conclude that;
\begin{equation}\begin{aligned}&
H\left(DV\right)= \frac{1}{2}\sigma_1^2 s_1^2 \frac{\partial^2 v}{\partial s_1^2}+ \frac{1}{2}\sigma_2^2 s_2^2 \frac{\partial^2 v}{\partial s_2^2}+...+\frac{1}{2}\sigma_n^2 s_n^2 \frac{\partial^2 v}{\partial s_n^2}+\\&\frac{\partial v}{\partial s_1}ds_1+\frac{\partial v}{\partial s_2}ds_2+...+\frac{\partial v}{\partial s_n}ds_n+
rs_1\frac{\partial v}{\partial s_1}+
rs_2\frac{\partial v}{\partial s_2}\\&+...+rs_n\frac{\partial v}{\partial s_n}+
\rho_{1n}\sigma_1 \sigma_n \frac{\partial^2 v}{\partial s_1s_n}-rv\end{aligned}
\end{equation}
where H(DV) being Hamiltonian interlinked with Black-Scholes-Merton. Here, our conclusion would be following
\begin{equation*}\begin{aligned}&
V_t+H\left(DV\right)=0 \quad where\quad V\in R^n ,\quad t\in \left(0, \infty\right),\\&
V=\left(g_1,g_2,g_3...g_n\right), in \left(t=0\right) \end{aligned}
\end{equation*}
\subsection{}
Initial condition (t=0), for Hamilton Jacobi
\begin{equation*}\begin{aligned}&
U=\left(g_1,g_2,g_3,...,g_n\right)\end{aligned}
\end{equation*}
Whereas Initial condition (t=0), for Black-Scholes-Merton for an european call option is
\begin{equation*}
V\left(s_1,s_2,s_3,...s_n,T\right)=\{max\left(max\left(s_1,s_2,s_3,...s_n\right)-K\right\}
\end{equation*}
\subsection{}
To relate the Black-scholes-Merton to Hamilton-Jacobi, In case of Black-Scholes when
\begin{equation*}
\left(s_1,s_2,s_3,...s_n\right)
\end{equation*}
denotes an n-underlying assets, whereas spatial variables are considered here as underlying assets or actives.
\subsection{}
By interlinking Hamiltonian of both equations. The dynamical aspects of Black-scholes-Merton allows to evaluate options, representing behaviour of options as physical objects.
\subsection{}
As showed above the multi-dimensional Black-Scholes-Merton is a dynamical system, possessing an associated energy, which preserves by both double convex functions, Hamilton and Lagrangian, Starting from Hamiltonian
\begin{equation*}
H=P_i\dot{x}-L\left(t,\dot{x}^j,\dot{x}^j\right),
\end{equation*}
\begin{equation*}
 \text{where}\quad P_i \quad \text{is the generalized momenta and taking}
\end{equation*}
\begin{equation*}\begin{aligned}&
P_i\dot{x}^i=\phi \quad \text{a potential function, and depicting}\end{aligned}
\end{equation*}
\begin{equation*}
L=\phi-H
\end{equation*}
Black-Scholes equation has lagrangian depicted by equation,
\begin{equation*}\begin{aligned}&
\mathbb{L}(s,\dot{s},t)=\phi-\{\frac{1}{2}\sigma_1^2 s_1^2 \frac{\partial^2 v}{\partial s_1^2}+ \frac{1}{2}\sigma_2^2 s_2^2 \frac{\partial^2 v}{\partial s_2^2}+...+\frac{1}{2}\sigma_n^2 s_n^2 \frac{\partial^2 v}{\partial s_n^2}\\&+\frac{\partial v}{\partial s_1}ds_1+\frac{\partial v}{\partial s_2}ds_2+...+\frac{\partial v}{\partial s_n}ds_n+
rs_1\frac{\partial v}{\partial s_1}+
rs_2\frac{\partial v}{\partial s_2}\\&+...+rs_n\frac{\partial v}{\partial s_n}+
\rho_{1n}\sigma_1 \sigma_n \frac{\partial^2 v}{\partial s_1s_n} \quad \text{with}\quad \phi=P\dot{s} \end{aligned}
\end{equation*}
\begin{equation*}
\text{By considering that}\quad\dot{s}\quad\text{ is the variation of the price of the action.}
\end{equation*}
\subsection{}
For a curious researcher it may be wealthy, to find methods which are applicable to problems in dynamics especially Hamilton-Jacobi method, to solve problems in financial Mathematics to understand the essence of interlinking of both equations.
\subsection{}
Multi-dimensional Black-scholes-Merton having solutions in the Metric of Banach space defined by,
\begin{equation*}
\left||V \right||=\left|V_1-V_2\right|
\end{equation*}
\begin{equation*}\begin{aligned}&
\text{whereas}\quad V_1\quad and\quad V_2\quad\text{ are solutions at time} \quad t_1\quad and\quad t_2\quad\\&
 \text{respectively}, \quad for\quad t_1,\end{aligned}
\end{equation*}
\begin{equation}\begin{aligned}&
V_1\left(s_1,s_2,...s_n\right)=s_1N\left(\phi_{11},p_1\right)+s_2N\left(\phi_{21},p_2\right)+\\&s_3N\left(\phi_{31},p_3\right)...+s_N\left(\phi_{n1},p_n\right)N
-Ke^{-rT}\{1-\{N\left(\phi_{11}'\right)+\\& N\left(\phi_{11}'\right)+N\left(\phi_{11}'\right)+...+N\left(\phi_{11}'\right)\}\}\end{aligned}
\end{equation}
For time $t_2$
\begin{equation}
\begin{aligned}&
V_2\{s_1,s_2,...s_n\}=s_1N\left(\phi_{12},p_1\right)+s_2N\left(\phi_{22},p_2\right)+\\&s_3N\left(\phi_{32},p_3\right)...+s_N\left(\phi_{n2},p_n\right)N
-Ke^{-rT}\{1-\{N\left(\phi_{12}'\right)+\\& N\left(\phi_{22}'\right)+N\left(\phi_{32}'\right)+...+N\left(\phi_{n2}'\right)\}\}\end{aligned}
\end{equation}
putting in above metric
\begin{equation}\begin{aligned}&
\sqrt{\left|v_1-v_2\right|}=\sqrt{s_1\left|N\left(\phi_{11}\right)-N\left(\phi_{12}\right)\right|+s_2\left|N\left(\phi_{21}\right)-N\left(\phi_{22}\right)\right|}+\\& \sqrt{s_3\left|N\left(\phi_{31}\right)-N\left(\phi_{32}\right)\right|
...+s_n\left|N\left(\phi_{n1}\right)-N\left(\phi_{n2}\right)\right|+Ke^{-rt}}\\& \sqrt{\{N\left(\phi_{11}'\right)-N(\phi_{12}')
+N\left(\phi_{21}'\right)-N\left(\phi_{22}'\right)+...+N\left(\phi_{n1}'\right)-N(\phi_{n2}')}\}\end{aligned}
\end{equation}
whereas $\sqrt{\left|v_1-v_2 \right|}=\sqrt{\triangle v}$\quad and
$N(\phi)=\int_{-\infty}^{\phi}e^{-w^2/2}dw$,
So we have $N(\phi_{11})-N( \phi_{21})$=\(\frac{1}{\sqrt{2\pi}}\)$\int_{-\infty}^{\phi_{11}}e^{-w^2/2}dw$-\(\frac{1}{\sqrt{2\pi}}\)$\int_{-\infty}^{\phi_{21}}e^{-w^2/2}dw$.
The similar procedure with integrals will continue until the nth difference in normal distribution functions,
\begin{equation*}
N(\phi_{n1})-N(\phi_{n2})= (\frac{1}{\sqrt{2\pi}}) \int_{-\infty}^{\phi_{n1}}e^{-w^2/2}dw-(\frac{1}{\sqrt{2\pi}})\int_{-\infty}^{\phi_{n2}}e^{-w^2/2}dw
\end{equation*}
\begin{equation}\begin{aligned}&
\sqrt{\left|v_1-v_2 \right|}=\sqrt{(\frac{s_1}{\sqrt{2\pi}})\int_{11}^{\phi_{12}}e^{-w^{2/2}}dw+(\frac{s_2}{\sqrt{2\pi}})\int_{21}^{\phi_{22}}e^{-w^{2/2}}dw+...+} \\&\sqrt{(\frac{s_n}{\sqrt{2\pi}})\int_{n1}^{\phi_{n2}}e^{-w^{2/2}}dw+1+(\frac{Ee^{-rT}}{\sqrt{2\pi}})\{\int_{-\infty}^{\phi_{11'}}e^{-w^2/2}dw+\int_{-\infty}^{\phi_{12'}}e^{-w^2/2}dw}\\&
\sqrt{+...+\int_{-\infty}^{\phi_{n1'}}e^{-w^2/2}dw+\int_{-\infty}^{\phi_{n2'}}e^{-w^2/2}dw\}}\end{aligned}
 \end{equation}
 After squaring both sides and using triangle inequality It becomes
 \begin{equation*}\begin{aligned}&
\left|v_1-v_2 \right|\leq(\frac{s_1}{\sqrt{2\pi}})\int_{11}^{\phi_{12}}e^{-w^{2/2}}dw+(\frac{s_2}{\sqrt{2\pi}})\int_{21}^{\phi_{22}}e^{-w^{2/2}}dw+...+ \\&(\frac{s_n}{\sqrt{2\pi}})\int_{n1}^{\phi_{n2}}e^{-w^{2/2}}dw+1+(\frac{Ee^{-rT}}{\sqrt{2\pi}})\{\int_{-\infty}^{\phi_{11'}}e^{-w^2/2}dw+\int_{-\infty}^{\phi_{12'}}e^{-w^2/2}dw\\&+...+\int_{-\infty}^{\phi_{n1'}}e^{-w^2/2}dw+\int_{-\infty}^{\phi_{n2'}}e^{-w^2/2}dw\}\end{aligned}
 \end{equation*}
 \subsection{}
 Now for the Lipschitz condition, to be verified by our initial condition,
 \begin{equation*}\begin{aligned}&
 V(s_1,s_2,...s_n,T)=max\{max(s_1,s_2,...s_n)-K,0\}
 \quad \\
 &choose \quad x_1,x_2,x_3,...,x_n ,\quad  y_1,y_2,y_3,...,y_n \in R^n\end{aligned}
 \end{equation*}
 \begin{equation*}\begin{aligned}&
\left|V\left(x_1,x_2,x_3,...,x_n,t\right)-V\left(y_1,y_2,y_3,...,y_n,t\right) \right|\\& \leq E\left|\left( x_1,x_2,x_3,...,x_n\right)-\left( y_1,y_2,y_3,...,y_n\right)\right|\end{aligned}
\end{equation*}
 \begin{equation*}
 For \quad\left(s_1,s_2,...s_n\right) \geq 0
 \end{equation*}
 \begin{equation*}\begin{aligned}&
 V(s_1,s_2,...s_n,T)= (s_1-s_2-s_3,...-s_n-K), for \quad  \left(s_1,s_2,...s_n\right) \geq K\\
 & and\quad0, \quad for  \quad \quad \left(s_1,s_2,...s_n\right) \leq K \end{aligned}
\end{equation*}
 \begin{equation*}\begin{aligned}&
\implies \left|V\left(x_1,x_2,x_3,...,x_n,T\right)-V\left(y_1,y_2,y_3,...,y_n,T\right) \right|\\&\leq E\left|\left( x_1,x_2,x_3,...,x_n\right)-\left( y_1,y_2,y_3,...,y_n\right)\right| \\
&  \quad for \quad \left(x_1,x_2,...,x_n\right)-\left(y_1,y_2,...y_n\right) \geq K\end{aligned}
\end{equation*}
\begin{equation*}
and \quad 0,\quad for    \quad \left(x_1,x_2,...,x_n\right)-\left(y_1,y_2,...y_n\right) \leq K
\end{equation*}
\begin{equation*}\begin{aligned}&
\left|V\left(x_1,x_2,x_3,...,x_n,t\right)-V\left(y_1,y_2,y_3,...,y_n,t\right) \right|\\&= \left|\left( x_1,x_2,x_3,...,x_n\right)-\left( y_1,y_2,y_3,...,y_n\right)\right|\end{aligned}
\end{equation*}
E=1, While V appears to be short map.
\section{conclusion}
We prove that Multi-dimensional Blac-scholes-Merton model is one of the dynamic systems. which is in this case Hamilton-Jacobi in n-variables  equivalent. we have found various relationships between both models by analyzing the solutions of Black-Scholes-Merton in suitable metric. Further studies could be conducted as an applications to solve financial problems by the methods of dynamics.

\end{document}